\documentclass{ws-ijmpa}

\begin{document}

\markboth{Makoto Oka}
{Spectroscopy of Pentaquark Baryons}

%
\catchline{}{}{}{}{}
%

\title{SPECTROSCOPY OF PENTAQUARK BARYONS%
\footnote{in collaboration with T.~Doi, H.~Iida, N.~Ishii, Y.~Nemoto, F.~Okiharu, H.~Suganuma and J.~Sugiyama.}
}

\author{\footnotesize MAKOTO OKA}

\address{Department of Physics, H27, Tokyo Institute of Technology\\
Meguro, Tokyo 152-8551, JAPAN\footnote{email: oka@th.phys.titech.ac.jp}
}

\maketitle

\pub{Received (Day Month Year)}{Revised (Day Month Year)}

\begin{abstract}
A review is given to pentaquark mass predictions in quark models and
QCD.  It is pointed out that no successful quark model prediction is
available for low-lying pentaquark states.
Some new results of direct application of QCD, QCD sum rules and lattice QCD, are also presented.

\end{abstract}

\section{Introduction}	
In 2003, the LEPS group at SPring-8 observed a sharp $\Theta^+$ peak in 
$\gamma n \to K^- \Theta^+$ reaction.\cite{Nakano}
This is an evidence of an exotic baryon resonance with strangeness $+1$, and
in the following year, it was confirmed by DIANA, CLAS, SAPHIR, HERMES, ZEUS, 
and COSY groups.\cite{Pos}
It is, however, challenged by multiple negative results by now, mostly from high energy experiments,
by the CDF, HyperCP, E690, BES, BELLE, BaBar, HERA-B, PHENIX groups.\cite{Pos}
This year new high statistics data by the CLAS collaboration at JLab have
revealed that a preceding observation of $\Theta^+$ by the same group was no more valid.\cite{CLASn}
The negative results have high statistics and are quite convincing, but
they may not completely wash away the ``evidence'' yet.
A new result from LEPS shows further evidence of 
forward photoproduction $\gamma+ d \to \Lambda(1520) +\Theta^+$.\cite{Nakano2}
The LEPS group claims that the result is not inconsistent with the other experimental data.
As it certainly requires confirmation, it will take some more time before the final conclusion.
Meanwhile, many theoretical works have been published.\cite{Theory}
A purpose of this article is to review the results in quark models
and also to present some new results from direct application of QCD.

$\Theta^+$ is considered to be a baryon with
$S = +1$, $B=1$, $Q=+1$ and $Y = B+S = 2$. 
Required ``valence quark'' combination is $uudd\bar s$ and thus
it is called pentaquark.
Except for a preliminary indication by the STAR group at RHIC,\cite{STAR} 
no $pK^+$ ($I_3= +1$) state is found, which suggests that 
the isospin of $\Theta^+$ is 0.
The simplest SU(3) irreducible representation is  $F =\overline{10}$, which is
unique within the flavor representations produced by five valence quarks.
The other members of the antidecuplet include genuine pentaquark states,
$\Xi_{I=3/2}$.

An often asked question is what we shall learn if such states exist.
In fact, QCD allows not only quark-anti-quark mesons and three-quark baryons, but
also various exotic states, {\it i.e.}, four-quark mesons, $qq\bar q \bar q$,
pentaquark baryons, $qqqq\bar q$, hybrid mesons, $\bar q g q$, 
dibaryons, $qqqqqq$, and so on.
Such exotic states will reveal multi-quark dynamics that cannot be accessed in 
non-exotic hadrons.
In particular, some important questions could be answered, such as
(1) whether the color confinement is simply bag-like, or string-like, or more complicated in multi-quark systems, 
(2) whether valence quark model is valid in hadronic interactions,
(3) what is the mechanism of dissociation of multi-quark states into two (or more) color-singlet hadrons,
and so on.
So far, the pentaquarks do not give definite answers to these questions, but it is clear that if $\Theta^+$ is a pentaquark state, then the standard constituent quark model should be modified so that a five quark state is
as light as the standard three quark baryons.

In this article, we first summarize current status of quark model calculations and then present some recent results from QCD sum rules and lattice QCD.

\section{Theory: I. Quark Models}
It is fair to say that there has been no successful calculation that fully explains $\Theta^+(1540)$.
First of all, most mass predictions are much higher than the observed mass, 1540 MeV.
Most calculations predict several low-lying excited states, but none is found.
For instance, many models predict low-lying $1/2^-$, $1/2^+$, $3/2^-$ and $3/2^+$ states ($I = 0$ and/or 1).

Several recent calculations try to find exact ground states of 5-quark (5Q) systems.
Such attempts include a molecular dynamics calculation, and variational calculations with and
without coupling to continuum states.
En'yo et al.\ applied an antisymmetrized molecular dynamics to five quarks and predicted several narrow states as the lowest energy states.\cite{Enyo}  
They showed that the intrinsic 5Q configuration prefers diquark clustering as was conjectured by Jaffe and Wilczek.\cite{JW} The absolute masses of the pentaquark states, however, had to be adjusted to the observed one and thus only the excitation energies can be predicted.

A variational calculation was performed by Takeuchi and Shimizu using one-gluon exchange plus
Nambu-Goldstone boson exchanges between quarks.\cite{Takeuchi}
They predicted several low-lying states within 200 MeV
above the ground state $1/2^-$, which they regard as a nonresonant state.  
But the lowest resonance state comes at above 2 GeV, that is about 500 MeV heavier than the desired pentaquark.
Hiyama et al.\ carried out a variational calculation of $1/2^-$ and $1/2^+$ pentaquarks 
with full couplings to KN continuum states.\cite{Hiyama}
They found that most of the low-lying states obtained in the bound-state approach (which constrains the wave function of five quarks to be localized) are dissolved into KN continuum.
Instead, they found a very narrow resonance state at about 500 MeV above the KN threshold.  Another highly-excited broad resonance was found in the spin $1/2^+$. Origins of these highly-excited resonances are not known yet.

It is clear now that the lowest $1/2^-$ state, which appears naturally in any quark model calculations, will not make a sharp resonance, but is almost completely mixed with KN scattering states.  Thus the next question is whether a $J = 1/2^+$ state is a possible candidate of narrow $\Theta^+$. 
It was shown by Hosaka et al.\ that it can be as narrow as a few MeV according to the quark structure of the p-wave pentaquark,\cite{Hosaka} 
but is always associated with a $J = 3/2^+$ state close-by. 
Their splitting is given by the LS force  and will be about a few tens MeV.
Because the partner is to have a similar narrow width, it should also be detected in the same production channel.

The next possibility is a $J = 3/2^-$ state, which is a viable candidate of $\Theta^+$, although mass predictions are still too high.
Because it decays only into d-wave KN scattering states, it is expected to be quite narrow. The spectroscopic factor is also highly suppressed, if we assume that the main component of the quark wave function consists of s-waves only.
In all, $1/2^+$, $3/2^+$ and $3/2^-$ may be candidates of the lowest resonance state of five quarks, but the predicted masses are too high to be identified with the observed $\Theta^+$.

\section{Theory: II. QCD Sum Rules}

It is highly desirable to apply QCD directly to the pentaquark problem under the situation.
We report here the results of two approaches in QCD.
The first one is to apply QCD sum rule to the pentaquarks of spin 1/2 and 3/2 with masses about  $1.5-2$GeV.  The second approach is lattice QCD (LQCD) in quenched approximation.

In both the approaches, one calculates a two-point correlation function of a local product of quark field operators.  This local operator is called interpolating
field operator (IFO).
We choose IFO for $J =1/2$ ($I=0$) pentaquark systems as
\begin{eqnarray}
J (x) &\equiv& 
\epsilon_{abc} \epsilon_{ade} \epsilon_{bfg} [u^T_d(x) C\gamma^5 d_e(x)][u^T_f(x) C d_g(x)] C \bar s^T_c(x) ,
\end{eqnarray}
and Rarita-Schwinger forms for	$J =3/2$ ($I=0$) pentaquark,
\begin{eqnarray}
J^{(1)}_{\mu}(x) &\equiv& 
\epsilon_{abc} \epsilon_{def} \epsilon_{cfg} [u^T_a(x) C d_b(x)][u^T_d(x) C\gamma_5\gamma_{\mu}  d_e(x)]
 \gamma_5 C \bar s^T_g(x) , 
 \label{eq:J1}\\
J^{(2)}_{\mu}(x) &\equiv& 
\epsilon_{abc} \epsilon_{def} \epsilon_{cfg} [u^T_a(x) C\gamma^5 d_b(x)]
[u^T_d(x) C\gamma_5\gamma_{\mu}  d_e(x)]
C \bar s^T_g(x) .
\label{eq:J2}
\end{eqnarray}
These operators are constructed from scalar, pseudoscalar and vector diquark combinations.  We also employ IFO's which are products of N and K or K$^*$ hadrons.  Although they are all local operators and are related by Fierz transform among them, each may have different coupling strengths to pentaquark resonance states, if they exist.

In the QCD sum rule,\cite{SDO} we parametrize the imaginary part of the correlation 
function in terms of a supposed resonance peak and continuum background
and determine the resonance parameters by comparing the correlation function 
with that calculated perturbatively at a large Euclid momentum.

We study whether the QCD spectral function for the relevant quantum numbers satisfies the positivity condition.
We have found that the correlation function of $J=1/2^+$ ($I=0$) pentaquark becomes almost zero or negative in the region of $M\sim 1-2$ GeV.
Thus it is concluded that there should be no resonance state in the same
mass region.
In contrast, $J = 1/2^-$ ($I=0$) pentaquark is not rejected in the same energy region.  Using the Borel sum rule technique, we determine its mass, although the result is sensitive to the chosen continuum threshold.  We find that the mass is around $1.3-1.7$ GeV.

Similarly, $J=3/2$ possibility is studied in the sum rules.\cite{Nishikawa,SDO2}
It is found that the sum rule depends sensitively on the choice of IFO, which indicates that the operators have different coupling strengths to a resonance state.  The second operator, $J^{(2)}_{\mu}$, that consists of a scalar and vector diquarks, happens to have a stronger strength at low energy.  On the other hand, $J^{(1)}_{\mu}$ does not give a positive spectral function in the $M\sim 1-2$ GeV region.

We have found that low lying $3/2^-$ and $3/2^+$ pentaquarks may exist from the positivity analysis and mass extraction from the $J^{(2)}_{\mu}$ sum rules.
In particular, the $3/2^-$ state is lower in mass, $\sim 1.5$ GeV, which is almost degenerate to $1/2^-$ state obtained in the sum rule.  Because the
$3/2^-$ state does not couple to s-wave KN scattering states, its spectral function is expected to be suppressed near the threshold.  It is reasonable to conclude that the $3/2^-$ pentaquark is a physical state at around 1.5 GeV.

\section{Theory: III. Lattice QCD}
Several lattice QCD calculations have been performed for the pentaquarks with $J=1/2$ and $J=3/2$.\cite{lattice}
They all take quenched approximation, while fermion formulations vary.
The conclusions are somewhat scattered at this moment in the sense that a few results indicate a pentaquark resonance state, while the majority denies existence of a low-lying sharp resonance.

We here present a new result in studies using an anisotropic lattice so that the precision of the mass estimate is high enough.\cite{Ishii}
We employ a lattice size $12^3\times 96$, and $\beta= 5.75$, which correspond to $(2.2)^3\times 4.4$ fm$^4$ in physical unit.
504 gauge configurations are generated on an anisotropic lattice ($a_s/a_t=4$), and the clover Wilson quark is employed.
Plots of effective masses show clear plateaus in most of the cases, which indicate a state with a definite energy is reached.
We have found that the diquark operators show less clear plateau structures than the operator constructed as a product of two hadrons, in particular for $J=3/2$. 

It is important to distinguish resonance states from hadronic scattering states, in the present case, mostly NK scatterings.
In order to judge whether a plateau state in an effective mass plot is a compact resonance state or not, we have invented a simple method in which the boundary conditions for some of the quarks are twisted.  For the $\Theta^+$ state, we perform a LQCD calculation where we impose antiperiodic boundary condition in the spatial directions on $u$ and $d$ quarks, keeping the $s$ quark in the periodic boundary condition (hybrid boundary condition).  Under such boundary conditions, pentaquark states, which have four compact $u+d$ quark contents, may not be affected much, while the nucleon and kaon, both of which contain odd number of $u+d$ quarks, should not be allowed to occupy the node-less ground state configuration.  Thus it is expected that the NK threshold energy will rise when the pentaquarks stay at the same masses. 

Using the hybrid boundary condition method, we have tested whether the obtained five quark plateaus are compact pentaquarks or not.
The results are unfortunately negative, i.e., all the plateaus are shifted according to the threshold so that they are all consistent with NK scattering states.
The results can be summarized as follows.
(1) A $J =1/2^-$ state observed at 1.75 GeV is likely to be a $L=0$ NK 
scattering state. The hybrid boundary condition method shows no compact 5Q states.
(2) A $1/2^+$ appears at 2.25 GeV, which is too heavy to be identified with $\Theta^+(1540)$.
(3) A $J =3/2^-$ state observed at 2.11 GeV on the lattice is consistent with 
NK$^*$ ($L=0$) scattering state.
(4) Similarly, $3/2^+$ states observed at 2.42 GeV and 2.64 GeV
are consistent with NK$^*$ ($L=1$) and N$^*$K$^*$ ($L=0$) scattering
states, respectively.

Thus no evidence of compact 5Q resonance state is obtained 
in the present LQCD calculation.
It should however be noted that the current calculation is under the
quenched approximation and also contains ambiguity due to extrapolation from the results for rather large quark masses.

\section{Conclusion}
In conclusion, we find no definite evidence of pentaquarks either in the 
quark model approaches or direct QCD calculations.
The quark model calculations of the 5Q mass have a difficulty 
that the predicted masses are still too high 
and also that the predictions are associated with some missing low-lying states.

Both the LQCD and QCD sum rule results reject existence of a low-lying $J =1/2^+$ 5Q state.
Although a low-lying $J =1/2^-$ state is predicted, it is found
that the state is not a compact resonance, but is likely to be a NK S-wave scattering state.
We have found that the hybrid boundary condition method is useful in determining
whether a low-lying state is a sharp compact resonance or a scattering state
of two hadrons.
The QCD sum rule allows $J =3/2^-$, and $3/2^+$ 5Q states, 
but the same states seen in LQCD are all consistent with two-hadron
(NK or N$^*$K scattering states.

This discrepancy between the results of the sum rules and lattice QCD on $J=3/2$ pentaquarks requires further study.  Both of the approaches have sone defects.  In the sum rule, convergence of the operator product expansion has to be checked. Because the pentaquark operators have higher dimensions than three-quark baryon operators, the pentaquark sum rules may suffer slower convergence.  The lattice QCD so far is under the quenched approximation.  It is known that the extrapolation in quark mass may cause large ambiguity in the quenched approximation.  It is highly desirable to perform unquenched lattice calculation with smaller quark masses.

\section*{Acknowledgments}

I acknowledge all the collaborators listed in the footnote of the headline.
I also acknowledge Drs.~S.~Takeuchi, T.~Shinozaki, T.~Nishikawa, A.~Hosaka and E.~Hiyama for discussions and information.

\end{document}